\documentclass[
  final,
  5p,
  12pt,
  twocolumn,
  number,
  sort&compress,
]{elsarticle}

\usepackage{amssymb}
\usepackage{amsmath}
\usepackage{graphicx}
\usepackage{tabularx}
\usepackage{booktabs}
\usepackage{subfig}
\usepackage{hyperref}
\hypersetup{colorlinks, citecolor=blue, linkcolor=blue,urlcolor=blue,}

\journal{\href{http://arxiv.org/abs/2307.14256}{arXiv:2307.14256}, published in \href{https://doi.org/10.1142/S0217732323502012}{Mod.Phys.Lett.A 39 (2024) 2350201.}}

\def\planck{E_\text{Planck}}

\def\gev{\text{GeV}}
\def\tev{\text{TeV}}
\def\grb{GRB 221009A}
\def\td{\text{d}}
\def\eliv{E_\text{LIV}}
\def\emax{E_\text{max}}

\def\fluxint{F_\text{int}}
\def\fluxobs{F_\text{obs}}
\def\fluxobsliv{F_\text{obs}^\text{LIV}}

\def\taugg{\tau_{\gamma\gamma}}
\def\tauggliv{\tau_{\gamma\gamma}^\text{LIV}}

\newcommand{\rw}{\(\mathbf{R_{WCDA}}\)}
\newcommand{\rk}{\(\mathbf{R_{KM2A}}\)}

\usepackage[normalem]{ulem}
\usepackage{xcolor}
\newcounter{num}
\ifodd \value{num}
{\gdef\new#1{{\color{blue}{#1}}}
\gdef\old#1{{\color{red}{\sout{#1}}}}}
\else
{\gdef\new#1{#1}
\gdef\old#1{}}
\fi

\usepackage[switch,mathlines]{lineno}
\let\oldequation\equation
\let\oldendequation\endequation
\renewenvironment{equation}
{\linenomath\oldequation}
{\oldendequation\endlinenomath}

\begin{document}

\begin{frontmatter}

    \title{Lorentz invariance violation from GRB 221009A}

    \author[inst1]{Hao Li}
    \ead{haolee@pku.edu.cn}

    \affiliation[inst1]{organization={School of Physics, Peking University},
        city={Beijing 100871},
        country={China}}

    \author[inst1,inst2,inst3]{Bo-Qiang Ma\texorpdfstring{\corref{cor1}}{}}
    \ead{mabq@pku.edu.cn}
    \cortext[cor1]{Corresponding author}

    \affiliation[inst2]{organization={Center for High Energy Physics, Peking University},
        city={Beijing 100871},
        country={China}}

    \affiliation[inst3]{organization={Collaborative Innovation Center of Quantum Matter},
        city={Beijing},
        country={China}}

\begin{abstract}
The Large High Altitude Air Shower Observatory~(LHAASO) reported observation of photons with energies above 10~TeV from gamma-ray burst GRB 221009A. A suggestion was proposed that this result may contradict our knowledge of special relativity~(SR) and the standard model~(SM), according to which photons of about 10~TeV from such a distant object should be severely suppressed because of the absorption by extragalactic background light. As a result, a number of mechanisms have been proposed to solve this potential puzzle, including Lorentz invariance violation~(LIV). In this work, we perform a detailed numerical calculation and show the feasibility 
to constrain LIV of photons from the LHAASO observation of GRB 221009A
quantitatively.
\end{abstract}


\end{frontmatter}

On October 9, 2022, an extremely bright (long) gamma-ray burst~(GRB), dubbed \grb\ was observed by several observatories, including the Fermi Gamma-ray Space Telescope~\cite{GBM1,GBM2,LAT1,LAT2} and the Large High Altitude Air Shower Observatory~(LHAASO)~\cite{LHAASO,lhaaso2023}.
As a long burst, \grb\ is rather close to the Earth, and \old{locates}\new{located} at \(z=0.1505\) in redshift~\cite{Redshift1,Redshift2}.
What makes this GRB deserve more attention is the observation of photons with energies up to about 18~TeV by LHAASO~\cite{LHAASO}.
In the standard understanding, the universe
is not always transparent to photons due to the background light absorption of high energy photons propagating in the space, and therefore it is natural to ask\old{y} the question: why such high energy photons from \grb\ \old{is}\new{are} permissible~\cite{Li:2022vgq,Li:2022wxc}.
Besides, another question follows that if within the standard framework of physics we cannot find an answer to this question, how to understand this phenomenon, and whether it is necessary to invoke new mechanisms (see, e.g., discussions and references in Ref.~\cite{Wang:2023okw}).

During the observation of LHAASO, the two different detectors, the Water Cherenkov Detector Array~(WCDA) and the Kilometer Square Array~(KM2A) both recorded \grb\, while at very different energy bands~\cite{Cao2010,LHAASO:2019qtb}.
WCDA reported more than 64000 photons with energies between 0.2~TeV and 7~TeV~\cite{lhaaso2023}, while the preliminary result from KM2A shows that there might be more than 5000 photons from 0.5~TeV to around 18~TeV\@~\cite{LHAASO}.
For convenience we may call the WCDA result as \rw\ and the KM2A report as \rk{}.

Recently a series of work has been devoted to analyzing possibility of excesses in the observation of \grb\ mainly based on \rk{}.
It seems that considering the absorption of extragalactic background light~(EBL), we could not observe about 18~TeV photons from \grb\ with LHAASO\@~\cite{Li:2022vgq,Li:2022wxc}.
Consequently it is suggested that there exists necessity for mechanisms beyond the special relativity~(SR) and the standard model~(SM) to explain the observed results.
Amongst the many attempts to understand this possible excess, we mainly focus on Lorentz invariance violation~(LIV) induced threshold anomalies~\cite{Li:2022vgq,Li:2022wxc} in this work below. \new{It is necessary to
mention that there are also studies~\cite{Zhu:2022usw,Finke:2022swf,Piran:2023xfg} to constrain LIV parameters by analyzing LIV induced time delays between photons with different energies, based on an alternative method to study the LIV effect.
Suggestions concerning non-standard mechanisms are also widely discussed, including utilizing axion-like particles~(ALPs)~\cite{Galanti:2022pbg,Baktash:2022gnf,Troitsky:2022xso,Nakagawa:2022wwm,Zhang:2022zbm,Wang:2023okw} and heavy/sterile neutrinos~\cite{Cheung:2022luv,Smirnov:2022suv,Brdar:2022rhc,Huang:2022udc,Guo:2023bpo} to understand the LHAASO observations.}


For our analysis, the relevant process in which LIV induced threshold anomalies take effect is the pair-production process \(\gamma\gamma_b\to e^-e^+\) with the background photons \(\gamma_b\) coming from EBL\@.
In the standard understanding of this process, the absorption can be calculated quantitatively by the optical depth~\cite{Biteau2015}:
\begin{equation}
\begin{aligned}
    \tau_{\gamma\gamma}(E,z) & =\int^z_0\td z^\prime\frac{\partial l}{\partial z^\prime}(z^\prime)\int_0^\infty\td\varepsilon\frac{\partial n}{\partial\varepsilon}(\varepsilon, z^\prime)\nonumber \\
                             & \times\int_{-1}^{+1}\td\mu\frac{1-\mu}{2}\sigma[\beta(E,z^\prime,\varepsilon,\mu)],\label{tau}
\end{aligned}
\end{equation}
where \(E\) is the energy of a GRB photon and \(z\) is the redshift of the GRB, \(\partial n/\partial\varepsilon(\varepsilon,z^\prime)\) is the number density of EBL photons of energy \(\varepsilon{}\) at redshift \(z^\prime{}\), and
\begin{equation}
    \frac{\partial l}{\partial z^\prime} = \frac{1}{H_0}\frac{1}{1+z^\prime}\frac{1}{\sqrt{\Omega_\Lambda+\Omega_M{(1+z^\prime)}^3}}
\end{equation}
is a factor from the \old{Fermi-Robertson-Walker}\new{Friedmann–Lemaître–Robertson–Walker metric of the standard model of} cosmology with \(H_0=70~\text{kms}^{-1}\text{Mpc}^{-1}\), \(\Omega_\Lambda=0.7\) and \(\Omega_M=0.3\).
In Eq.~\eqref{tau} the pair-production process cross-section is~\cite{Nikishov1962,Gould1967b}:
\begin{equation}
\begin{aligned}
    \sigma(\beta) & =\frac{3\sigma_T}{16}(1-\beta^2)\nonumber                                                        \\
                  & \times\left[2\beta(\beta^2-2)+(3-\beta^4)\ln\frac{1+\beta}{1-\beta}\right],\label{cross-section}
\end{aligned}
\end{equation}
where \(\sigma_T\) is the Thomson cross-section and
\begin{equation}
    \beta(E,z,\varepsilon,\mu)=\sqrt{1-\frac{2m_e^2}{E\varepsilon}\frac{1}{1-\mu}{\left(\frac{1}{1+z}\right)}^2}.\label{beta}
\end{equation}
It is noteworthy that Eq.~\eqref{beta} should be real so that the threshold information in SR is already encoded and hence the integration in Eq.~\eqref{tau} is understood to be performed satisfying the condition~\(\beta\in\mathbb{R}\), which indicates the ordinary threshold condition:
\begin{equation}
    E_{thr}=\frac{2m_e^2}{\varepsilon}\frac{1}{1-\cos\theta}\left(\frac{1}{1+z}\right){}^2.\label{thres}
\end{equation}

However, in certain models, LIV could cause threshold anomalies in various reactions, including the pair-production process~\cite{LI2021a}.
The threshold anomaly in the pair-production process is suggested to modify the threshold condition in Eq.~\eqref{thres}, and it is possible that the \(\varepsilon{}\)-space integrated in Eq.~\eqref{tau} gets shrunk, resulting in a smaller optical depth \(\tauggliv{}\).
By considering the intrinsic flux~\(\fluxint{}\), the observed flux~\(\fluxobs{}\) and the observed flux with LIV~\(\fluxobsliv{}\), which are related by
\begin{equation}
\begin{aligned}
    F^\text{LIV}_\text{obs} & :=F_\text{int}\times e^{-\tau^\text{LIV}_{\gamma\gamma}},\nonumber \\
    F_\text{obs}            & :=F_\text{int}\times e^{-\tau_{\gamma\gamma}},
\end{aligned}
\end{equation}
from which we may have
\begin{equation}
    F_\text{obs}^\text{LIV} > F_\text{obs},
\end{equation}
As a result, with a proper LIV parameter \(\xi{}\) we might be able to interpret \rk{}.
Here the parameter \(\xi{}\) is defined to be the deviation from the photon dispersion relation:
\begin{equation}
    \omega(k){}^2\approx k^2-\xi k^3+\cdots\label{mdr}
\end{equation}
with the photon energy \(\omega{}\) not too large compared to \(\planck{}\).
For later convenience, we also define \(\xi^{-1} \equiv \eliv{}\).

Now we concentrate on \grb{}, especially on the results \rw\ and \rk\ from LHAASO\@.
First we consider what we can learn from \rw{}.
For convenience we assume that there are exactly \(N_\text{WCDA} = 64000\) photons in total in this data set.
The distribution of these photons is assumed to follow a simple power-law form:
\begin{equation}
    \frac{\td N}{\td E}=A_\alpha\times E{}^{-\alpha},\label{spectrum}
\end{equation}
where the index \(\alpha{}\) is chosen to be \(2.41^{+0.14}_{-0.13}\)~\footnote{The values are taken from Ref.~\cite{lhaaso2023} and its supplementary material, where time-dependent values \(\alpha(t)\) are given and here we use the time-averaged values.} and \(A_\alpha{}\) is left to be normalized by solving
\begin{equation}
    N_\text{WCDA}=\int_{200~\gev }^{7~\tev } \td E\, \frac{\td N}{\td E}\times e^{-\tau},
\end{equation}
in which \(\tau{}\) could be either \(\taugg{}\) or \(\tauggliv{}\).
Once the normalization factor \(A_\alpha{}\) is determined for each case, we may extend the spectrum~\eqref{spectrum} to a higher energy \(\emax{}\) and thus we can compare the theoretical calculation with \rk{}.
The results of this comparison then would provide us with more clue of whether there is any unexpected excess in the observation of \grb{}.

In this work, we choose \(\emax=20~\tev \) and to calculate the optical depth, we adopt the model of Ref.~\cite{Dominguez2011}. 
Let us define a critical energy~\(E_c\), which satisfies
\begin{equation}
    \int_{E_c}^{\emax}\td E\,\frac{\td N}{\td E}\times e^{-\tau}=1,\label{ec}
\end{equation}
which means that there should be at least one 
event within ${E_c}$ to ${\emax}$.
By solving \(E_c\) we could obtain the information about at most above which energy can we observe one photon.
If \(E_c\gtrapprox 18~\tev{}\), then observing photons around 18~TeV is quite natural.
However when \(E_c\ll 18~\tev{}\), it is unlikely that we can do the same thing conclusively.
Replacing \(\taugg{}\) with \(\tauggliv{}\), we can perform the similar analyses with the existence of LIV\@.
The results then could indicate whether we need to invoke new mechanisms, and if so, whether LIV is a good candidate.

\begin{figure}
    \centering
    \subfloat[][]{\label{fig1a}\includegraphics[scale=0.54]{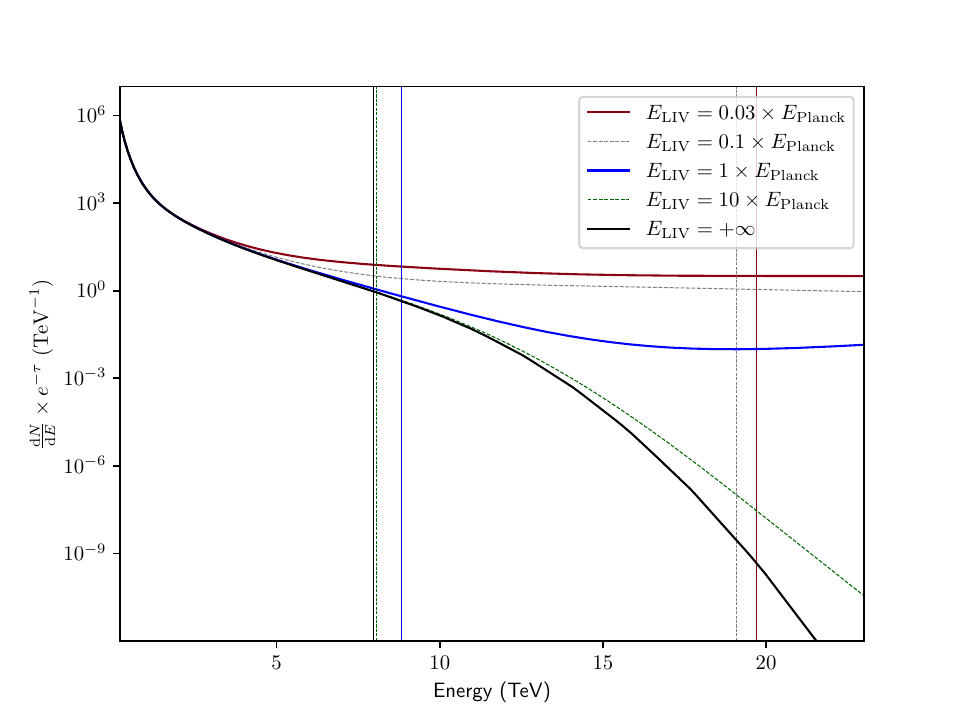}}\\
    \subfloat[][]{\label{fig1b}\includegraphics[scale=0.54]{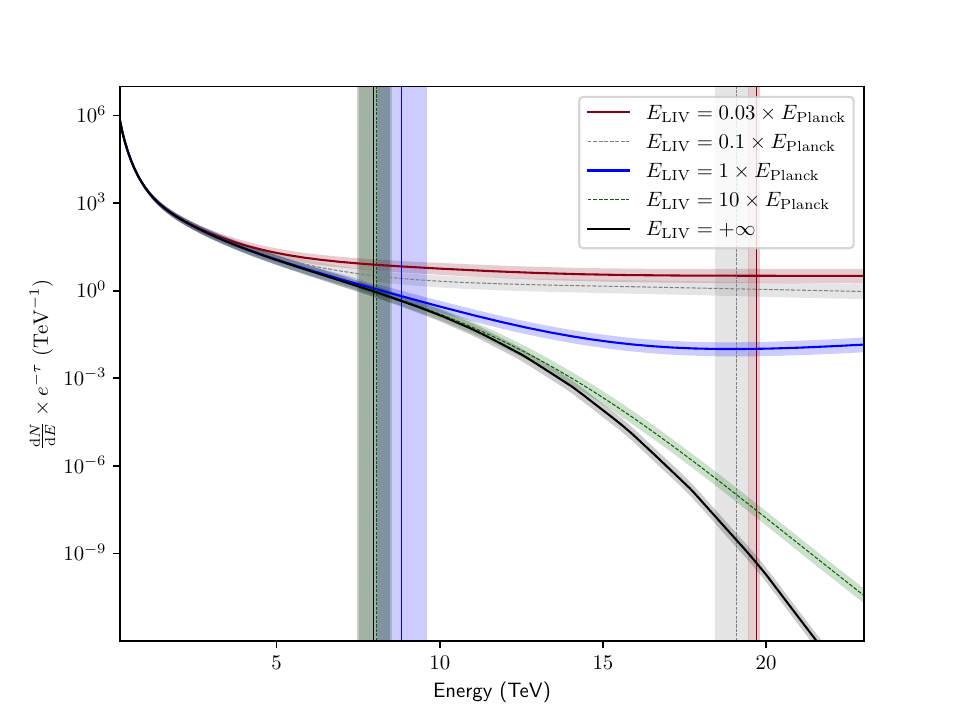}}%
    \caption[The expected spectra with different LIV scales.]{The expected spectra with different LIV scales: \subref{fig1a} with \(\alpha=2.41\) and the vertical lines representing the positions of \(E_c\); \subref{fig1b} the same as in~\subref{fig1a} with the filled regions representing the uncertainties.\label{fig1}}
\end{figure}

We choose the LIV scales to be \(\eliv=0.03\times\planck{}\)~\footnote{This is a phenomenological suggestion of the lower bound of \(\eliv{}\), see, e.g., 
Ref.~\cite{Zhu:2022usw} 
and references therein.}, \(\eliv=0.1\times\planck{}\), \(\eliv=\planck{}\), \(\eliv=10\times\planck{}\), and \(\eliv=+\infty{}\) which can also be understood as the non-LIV case.
In the case of \(\alpha=2.41\), we list the critical energies and the total photon numbers for the extended spectra in Tab.~\ref{tab1}, and meanwhile we depict the results in Fig.~\ref{fig1a}, where the spectra are drawn and the positions of \(E_c\) are shown by vertical lines schematically.
From Tab.~\ref{tab1}, we learn that the photon numbers only change mildly, since the suppression caused by EBL attenuation and the shape of the spectrum turns stronger at high energies, and as a result most photons still `{\it accumulate}' at low energies.
Then we concentrate on the positions of \(E_c\) for different LIV scales.
Obviously from the last three rows of Tab.~\ref{tab1} we may conclude that it is quite unlikely that LHAASO could report photons around 18~TeV from \grb{}, since there is only one photon above \(\sim 8~\tev{}\) and because of the aforementioned `{\it accumulation}' it is not consistent with \rk{}.
Indeed if we replace with \(10^{-6}\) the one in Eq.~\eqref{ec} and set \(\eliv=+\infty{}\), we are able to obtain the corresponding \(E_c\) in the second row of Tab.~\ref{tab4}.
It provides a more strict limit that there is no more than \(10^{-6}\) photon above 16.54~TeV, making it rather unusual to observe photons about 18~TeV\@.
However from the first two rows of Tab.~\ref{tab1} we find that a LIV scale \(\eliv\lesssim 0.1\times\planck{}\) is enough for understanding \rk{}, that is, there are photons observable around 18~TeV\@.

To minimize the effects caused by choosing different spectrum indices, we also perform the same analyses for \(\alpha=2.28\) and \(\alpha=2.55\), for which the results are depicted in Tab.~\ref{tab2} and Tab.~\ref{tab3} respectively.
As we in Fig.~\ref{fig1b} we depict the effects of changing the index schematically by the filled regions.
Meanwhile, in Tab.~\ref{tab4} the \(E_c\) for finding \(10^{-6}\) photon are also listed.
Needless to say, although with a larger index \(\alpha=2.55\) it would be more unlikely to observe photons around 18~TeV, LIV induced threshold anomalies still provide an explanation for \rk\ so long as \(\eliv\lesssim 0.1\times\planck{}\).
However even a smaller index \(\alpha=2.28\) could not provide a reasonable interpretation to \rk{}.
As we can see from Tab.~\ref{tab2} and Fig.~\ref{fig1b}, without LIV we might have only one photon above 8.46~TeV, while above 16.76~TeV only \(10^{-6}\) photon is observable.
In contrast, with LIV, or more precisely, with \(\eliv\lesssim 0.1\times\planck{}\), \rk\ from LHAASO can be understood naturally.

\new{
Other uncertainties may also originate from the selection of the model of EBL, and have already been analyzed in Ref.~\cite{Li:2023rgc}.
Here we give a brief description.
Indeed the model we adopted~\cite{Dominguez2011} captures the main characteristics of other 
well-known
models, and the qualitative properties of the results are universal.
Therefore the results would not have any difference in the orders of magnitude, and even the EBL model we adopted is replaced by other models, the analyses in this work are still valid.
}

\begin{table}[htbp]
    \centering
    \caption{\(E_c\) and expected photon numbers with \(\alpha=2.41\).\label{tab1}}
    \begin{tabular}{ccc}
        \toprule\toprule
        \(\eliv~(\planck)\) & \(E_c~(\tev)\) & Photon number \\
        \midrule
        0.03                & 19.69     & 64060     \\
        0.1                 & 19.01       & 64023   \\
        1                   & 8.82         & 64003  \\
        10                  & 8.05         & 64003  \\
        \(\infty{}\)        & 7.97          & 64002 \\
        \bottomrule\bottomrule
    \end{tabular}
\end{table}
\begin{table}[htbp]
    \centering
    \caption{\(E_c\) and expected photon numbers with \(\alpha=2.28\).\label{tab2}}
    \begin{tabular}{ccc}
        \toprule\toprule
        \(\eliv~(\planck)\) & \(E_c~(\tev)\) & Photon number \\
        \midrule
        0.03                & 19.81     & 64098     \\
        0.1                 & 19.48       & 64038   \\
        1                   & 9.61         & 64005  \\
        10                  & 8.55         & 64004  \\
        \(\infty{}\)        & 8.46          & 64004 \\
        \bottomrule\bottomrule
    \end{tabular}
\end{table}
\begin{table}[htbp]
    \centering
    \caption{\(E_c\) and expected photon numbers with \(\alpha=2.55\).\label{tab3}}
    \begin{tabular}{ccc}
        \toprule\toprule
        \(\eliv~(\planck)\) & \(E_c~(\tev)\) & Photon number \\
        \midrule
        0.03                & 19.45     & 64036     \\
        0.1                 & 18.43       & 64014   \\
        1                   & 8.09         & 64002  \\
        10                  & 7.53         & 64002  \\
        \(\infty{}\)        & 7.47          & 64002 \\
        \bottomrule\bottomrule
    \end{tabular}
\end{table}

\begin{table}[htbp]
    \centering
    \caption{\(E_c\) for observing \(10^{-6}\) photon without LIV for different indices.\label{tab4}}
    \begin{tabular}{p{3.5cm}<{\centering}p{3.5cm}<{\centering}}
        \toprule\toprule
        \(\alpha{}\) & \(E_c~(\tev)\) \\
        \midrule
        2.28 & 16.76 \\
        2.41 & 16.54 \\
        2.55 & 16.31 \\
        \bottomrule\bottomrule
    \end{tabular}
\end{table}

\new{
Besides the observations of LHAASO, it is also noteworthy that Carpet-2 reported a 251~TeV photon possibly from \grb{}~\cite{Carpet}.
Since the sensitivity of Carpet-2 is comparable to that of LHAASO, what we can infer from this work is that in the standard case, even a photon of about 20~TeV is not observable, therefore a 251~TeV photon from \grb\ is not explicable for Carpet-2.
Thus from a standard viewpoint, the Carpet-2 observation of a 251~TeV is not consistent with the observations of LHAASO.
However, new physics effects might make this observation possible, and we would like to demonstrate this in the following.
Let us assume that the LIV effect~(or other scenarios such as axion-like particles) is maximal, such that no photon from \grb\ would be absorbed during its propagation.
In other words, the spectrum we may observe is exactly the form of Eq.~\eqref{spectrum}.
Following the same logic of this work, and extending \(\emax{}\) to 300~TeV, we are able to calculate the photon number \(N_\text{exp}\) between 200~TeV and 300~TeV.
Here the index \(\alpha{}\) plays the role of a variable.
For \(\alpha=2.28\), \(N_\text{exp}=3.79\), for \(\alpha=2.41\), \(N_\text{exp}=1.65\), and for \(\alpha=2.55\), \(N_\text{exp}=0.67\).
Furthermore, to make \(N_\text{exp}=1\) we have \(\alpha=2.49\).
However, the index of a higher energy band is normally larger than that of a lower energy band, hence we find that, for LHAASO, there is no more than a couple of photons around 250~TeV even the LIV~(or other) effects are considered to be maximal.
Indeed, for \(\alpha=3\) we have \(N_\text{exp}=0.04\) and \(\alpha=5\) we have \(N_\text{exp}=5\times 10^{-8}\).
Taking into consideration the fact that the sensitivity for detecting photons of Carpet-2 is comparable to that of LHAASO, it suggests that only when the LIV~(or other novel mechanisms) effects are considerable can we interpret the observation of Carpet-2, although there do exist such options.
As an alternative explanation, it was suggested that the origin of this photon could be galactic~\cite{Carpet,Carpet2}.
We hope that future release of the Carpet-2 data would shed light on this problem, and here we only present the possibilities.
}

In summary, we perform analyses of the observation results of \grb\ by LHAASO~\cite{LHAASO,lhaaso2023}.
We utilize the detailed result, which exhibit no excess and no contradiction with the standard physics, from WCDA of LHAASO~\cite{lhaaso2023}, to construct models and extend these models to the preliminary results from KM2A of LHAASO~\cite{LHAASO}, which is likely to contradict special relativity.
After comparing the theoretical predictions with the observation, we find that 
around 18~TeV photons from \grb\ by LHAASO 
are hardly understood from standard physics, and novel mechanisms are needed to provide reasonable explanations.
Furthermore we explore the possibility of using Lorentz invariance violation to understand this observation.
We find that a LIV scale \(\eliv\lesssim 0.1\times\planck{}\) is enough to render about 18~TeV photons from \grb\ detectable on the Earth.
As a result, 10~TeV scale 
photons from \grb\ 
indicate
the feasibility to constrain LIV of photons from LHAASO observation.
Of course, besides LIV, there are also other approaches available, such as introducing axion-like particles~\cite{Baktash:2022gnf,Galanti:2022pbg,Troitsky:2022xso,Nakagawa:2022wwm,Zhang:2022zbm,Wang:2023okw} \new{and heavy/sterile neutrinos~\cite{Cheung:2022luv,Smirnov:2022suv,Brdar:2022rhc,Huang:2022udc,Guo:2023bpo}}.
Therefore we expect that more detailed analyses on the results of KM2A observation from LHAASO could provide more information.

\section*{Declaration of competing interest}

The authors declare that they have no known competing financial interests or personal relationships that could have appeared to influence the work reported in this paper.



\section*{Acknowledgements}

This work is supported by National Natural Science Foundation of China~(Grants No. 12335006 and No. 12075003).

\bibliographystyle{elsarticle-num}
\bibliography{ref}

\end{document}